\documentstyle[aps,prl,psfig]{revtex}
\begin{document}

\twocolumn[\hsize\textwidth\columnwidth\hsize\csname
@twocolumnfalse\endcsname
\title{Data clustering and noise undressing of correlation matrices}
\author{Matteo Marsili}
\address{Istituto Nazionale per la Fisica della Materia (INFM),
Trieste-SISSA Unit, V. Beirut 2-4, Trieste I-34014}
\date{\today}
\maketitle
\widetext

\begin{abstract}
We discuss a new approach to data clustering. We find that maximum 
likelyhood leads naturally to an Hamiltonian of Potts variables which 
depends on the correlation matrix and whose low temperature
behavior describes the correlation structure of the data. 
For random, uncorrelated data sets no correlation structure emerges. On
the other hand for data sets with a built-in cluster structure, the method
is able to detect and recover efficiently that structure. Finally we apply the
method to financial time series, where the low temperature behavior reveals 
a non trivial clustering. 
\end{abstract}

]
\narrowtext

Statistical mechanics typically addresses the question of how structures and 
order arising from interactions in extended systems are dressed, and 
eventually destroyed, by stochastic -- so-called thermal -- fluctuations. 
The inverse problem, unraveling the structure of correlations from stochastic 
fluctuations in large data sets, has recently been addressed using  
ideas of statisical mechanics\cite{Rose,Domany}. This is the case of
data clustering problems, where the goal is to classify $N$ objects, defined 
by $D$ dimensional vectors $\{\vec \xi_i\}_{i=1}^N$, in equivalence classes.
Generally\cite{Rose} the idea is {\em 1)} postulate a cost function, which 
depends on the data sample, for each structure and {\em 2)} consider the 
cost function as an Hamiltonian and study its thermal properties.
Structures are identified by configurations ${\cal S}=\{s_i\}_{i=1}^N$ of 
class indices, where $s_i$ is the equivalence class to which object $i$ 
belongs. 
Regarding $s_i$ as Potts spins, a Potts Hamiltonian $H_q=\sum_{i<j} 
J_{i,j}\delta_{s_i,s_j}$ has been recently proposed \cite{Domany} 
as a cost function, with couplings $J_{i,j}$ decreasing with the distance 
$d_{i,j}=||\vec \xi_i-\vec \xi_j||$ between objects $i$ and $j$.
The underlying structure of data sets emerges as the clustering of Potts
variables at low temperatures. 

In this work we address the question of data clustering for 
time series. Rather than postulating the form of the Hamiltonian, 
we start from a statistical {\em ansatz} and invoke maximum likelyhood
and maximum entropy principles. 
In this way, the structure of the Hamiltonian 
arises naturally from the statistical {\em 
ansatz}, without the need of assumptions on its form.
We study, by Montecarlo method, this Hamiltonian for artificial time series:
If time series are generated with some cluster structure ${\cal S}^\star$, 
we find a low temperature phase which is dominated by cluster configurations 
close to ${\cal S}^\star$. For random time series no low temperature phase is
found. We also study time series of assets composing the S\&P500 index, 
whose correlations have been the subject of much recent 
interest\cite{focus,Mantegna,KKM}. 
Correlation matrices of financial time series are of great practical 
interest. Indeed they are at the basis of risk minimization in the modern
portfolio theory\cite{CAPM}. This states that, in order to reduce
risk, the investment needs to be {\em diversified} (i.e. divided) on many
uncorrelated assets. However the measure of correlation in finite samples 
was recently found to be affected by considerable 
noise-dressing\cite{focus}. 

Our aim is to address the problem of revealing the structure of {\em bare} 
correlations hidden in a finite data set.
Quite interestingly, our analysis of the S\&P500 data set reveals a low 
temperature behavior dominated by few clusters of correlated 
assets with scale invariant properties. 
The thermal average over the relevant cluster structures provides
a good fit of the financial correlations, which allows us to estimate
the {\em noise-undressed} correlation matrix.
Finally, we discuss several generalizations of our approach to generic
data clustering.

\begin{figure}
\centerline{\psfig{file=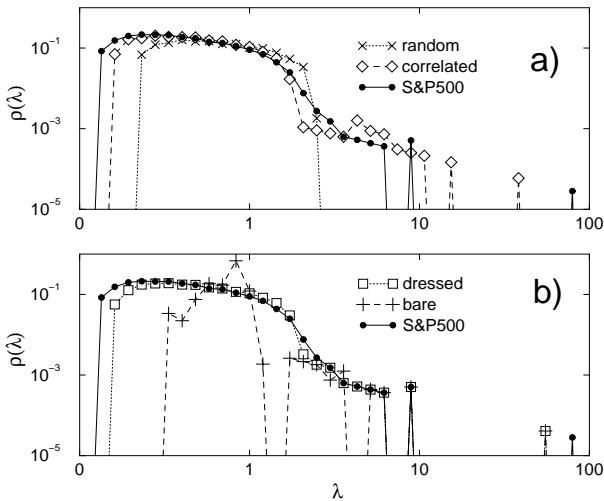,width=8cm}}
\caption{
{\bf a)} Distribution of eigenvalues of the correlation matrices 
of S\&P500 (full line $\bullet$) random (dotted $\times$) 
and correlated (dashed $\diamond$) time series. 
{\bf b)} Comparison of the spectrum of the S\&P500 correlation matrix
(full line $\bullet$) with noise-dressed (dotted $\Box$) 
and bare (dashed $+$) correlation matrices generated by Eq. 
(\ref{fitbeta}).}
\label{figmat}
\end{figure}

The data $\Xi=\{\vec\xi_i\}_{i=1}^N$ is composed of $N$ sets 
$\vec\xi_i=\{\xi_i(d)\}_{d=1}^D$
of $D$ measures. These are normalized to zero mean $\sum_d \xi_i(d)/D=0$
and unit variance $\sum_d \xi_i^2(d)/D=1$.
We focus below on the case where $\xi_i(d)$ is the normalized daily 
returns of asset $i$ of the S\&P500 index, in day $d$\cite{notanorma}. 
For the moment being, let us assume that $\xi_i(d)$ are Gaussian variables. 
The reason is that we want to focus exclusively on pairwise correlations 
and the Gaussian model is 
the only one which is completely specified at this level. We shall discuss
below how to apply the method when $\xi_i(d)$ are not Gaussian.
The key quantity of interest is the matrix
\begin{equation}\FL
C_{i,j}(D)\equiv \frac{1}{D}\sum_{d=1}^D\xi_i(d)\xi_j(d).
\label{corr}
\end{equation}
The spectral properties of this matrix, for uncorrelated time series,
are known exactly\cite{Sengupta}. The spectrum of eigenvalues $\lambda$
extends over an interval of size $\sim N/D$ around $\lambda=1$, as 
shown in Fig. \ref{figmat}a. The spectrum of eigenvalues of the S\&P500
correlation matrix is also shown. The similarity of the two distributions
for $\lambda\approx 1$ suggests that significant noise-dressing due to 
finite $D$ occurs\cite{focus}. The tail of the distribution ($\lambda\gg 1$) 
implies that some correlation is however present. The structure of 
correlation was analyzed both by minimal spanning tree method \cite{Mantegna}
and by the method of ref. \cite{Domany} in ref. \cite{KKM}. 

In order to explain this correlation, Noh \cite{Noh} proposed the 
{\em ansatz}
\begin{equation}\FL
\xi_i(d)=\frac{\sqrt{g_{s_i}}\eta_{s_i}(d)+\epsilon_{i}(d)}
{\sqrt{1+g_{s_i}}}.
\label{ansatz}
\end{equation}
Here $g_s>0$ and
 $s_i$ are integer variables (so-called Potts
spins), $\eta_s(d)$ and $\epsilon_i(d)$ are {\em iid} gaussian variables
with zero average and unit variance. 
In order to allow for totally uncorrelated sets, we allow
$s_i$ to take all integer values up to $N$. 
In Eq. (\ref{ansatz}) sets are correlated in clusters labeled by 
$s$. The $s^{\rm th}$ cluster is composed of $n_s$ sets with {\em 
internal correlation} $c_s$, where
\begin{equation}\FL
n_s=\sum_{i=1}^N \delta_{s_i,s},\qquad
c_s=
\sum_{i,j=1}^N
C_{i,j}\delta_{s_i,s}\delta_{s_j,s}.
\label{clsvar}
\end{equation}
The correlation matrix generated by Eq. (\ref{ansatz})
for $D\to\infty$ is 
$C_{i,j}=(g_{s_i}\delta_{s_i,s_j}+\delta_{i,j})/(1+g_{s_i})$.
Its distribution of eigenvalues is simple: 
To each $s$ with $n_s\ge 1$ there correspond one eigenvalue
$\lambda_{s,0}=(1+g_s n_s)/(1+g_s)$ 
and $n_s -1$
eigenvalues $\lambda_{s,1}=1/(1+g_s)$ 
Hence, large eigenvalues correspond to groups of many ($n_s\gg 1$)
sets. For $D$ finite, we expect noise to lift degeneracies 
between $\lambda_{s,1}$ but to leave the structure of large eigenvalues 
unchanged.

In order to fit the data set $\Xi$ with Eq. (\ref{ansatz}), let us 
compute the likelyhood. This is the probability $P(\Xi|{\cal S},{\cal G})$
of observing the data $\Xi$ as a realisation of Eq. (\ref{ansatz}) with 
structure ${\cal S}$ and 
parameters ${\cal G}=\{g_s\}_{s=1}^N$, and it reads
\[
P(\Xi|{\cal S},{\cal G})=\prod_{i=1}^N\prod_{d=1}^D
\left\langle
\delta\left(
\xi_i(d)-\frac{\sqrt{g_{s_i}}\eta_{s_i}(d)+\epsilon_{i}(d)}
{\sqrt{1+g_{s_i}}}\right)
\right\rangle
\]
where the average is over all the $\eta$'s and $\epsilon$'s variables
and $\delta(x)$ is Dirac's delta function.
Gaussian integration leads to $P(\Xi|{\cal S},{\cal G})
\propto e^{-DH\{{\cal S},{\cal G}\}}$ with $H\{{\cal S},
{\cal G}\}=\frac{1}{2}\sum_s[(1+g_s)
(n_s-\frac{g_s c_s}{1+ g_s n_s})-n_s\ln(1+g_s)
+\ln(1+ g_s n_s)]$. We fix 
the coupling strengths $g_s$ by likelyhood maximization 
$\frac{\partial H}{\partial g_s}=0$ for all $s$, which yields
\begin{equation}\FL
\hat g_s=\frac{c_s -n_s}{n_s^2-c_s}
\label{xhat}
\end{equation}
for $n_s>1$ and $\hat g_s=0$ for $n_s\le 1$.
Note that for uncorrelated sets
$C_{i,j}=\delta_{i,j}$ we have $c_s=n_s$ $\forall s$ and hence
$\hat g_s=0$. The coupling strength $\hat g_s$ instead diverges for totally
correlated sets ($C_{i,j}=1$) because
$c_s=n_s^2$. Using Eq. (\ref{xhat}) we find that the likelyhood of 
structure ${\cal S}$ under ansatz (\ref{ansatz}) takes the form 
$P(\Xi|{\cal S})\propto e^{-D H_c}$, where 
\begin{equation}\FL
H_c\{{\cal S}\}=\frac{1}{2}\sum_{s: n_s>0}
\left[\log
\frac{c_s}{n_s}+(n_s-1)\log\frac{n_s^2-c_s}
{n_s^2-n_s}\right]
\end{equation}
The ground state ${\cal S}_0$ of $H_c$ yields the maximum likelyhood
fit with Eq. (\ref{ansatz}). This would probably take the ansatz (\ref{ansatz})
too seriously. In general, it is preferrable to consider probabilistic
solutions $P\{{\cal S}\}$ and, following ref. \cite{Rose}, we invoke the
the maximum entropy principle: Among all distributions $P\{{\cal S}\}$
with the same average log-likelyhood, we select that which has maximal
entropy. This, as usual, leads to the Gibbs distribution 
$P\{{\cal S}\}\propto e^{-\beta H_c\{{\cal S}\}}$
where the inverse temperature $\beta$ arises as a Lagrange multiplier.

The Hamiltonian $H_c$ depends implicitely on the Potts spins $s_i$ 
through the cluster variables $n_s$ and $c_s$ of Eq. (\ref{clsvar}). 
Unlike the Potts Hamiltonian $H_q$, the dependence on $\delta_{s_i,s_j}$ is 
non-linear and it is modulated by $C_{i,j}$.
For $s_i\ne s_j$ for all $i\ne j$ we have $n_s=c_s=1$ for all $s$ and
hence $H_c=0$. This state is representative of the
high temperature ($\beta\to 0$) limit. The low temperature physics of
$H_c$ is instead non-trivially related to the correlation matrix
$C_{i,j}$. Note, that the ferromagnetic state
$s_i=1, \forall i$, which dominates as $\beta\to\infty$ in clustering methods 
based on Potts models\cite{Domany}, is in general not the ground state
of $H_c$. Intuitively we expect that, if the model of Eq. (\ref{ansatz}) 
is reasonable,
$H_c$ should have a well defined ground state and low temperature phase
which is energetically dominated by this state. In these cases, as in 
ref. \cite{Domany}, we expect a thermal phase transition\cite{notaglass}. 

In order to study the properties of $H_c$ we resort to
Montecarlo (MC) method by Metropolis algorithm\cite{Metropolis}.
This, at equilibration, allows us to sample the
Gibbs distribution $P\{{\cal S}\}$ and compute 
average quantities, such as the internal energy 
$E_\beta=\langle{H_c}\rangle_\beta$ where $\langle{\ldots}\rangle_\beta$ 
stands for thermal average.
In order to detect the occurrence of spontaneous magnetization --
which occurs if the $s_i$ lock into energetically favourable
configurations at low temperature -- 
we measure the autocorrelation function
\begin{equation}\FL
\chi(t,\tau)=\frac{\sum_{i<j}\delta_{s_i(t),s_j(t)}
\delta_{s_i(t+\tau),s_j(t+\tau)}}{\sum_{i<j}
\delta_{s_i(t),s_j(t)}}.
\label{chit}
\end{equation}\FL
This quantity tells us how many pairs of sites belonging to the same
cluster at time $t$ are still found in the same cluster after $\tau$ MC steps.
For $t$ large enough, $\chi$ becomes a function of $\tau$ only. 
This function decreases rapidly to a plateau value 
$\chi_\beta=\langle{\chi(t,\tau)}\rangle_\beta$ for $t\gg \tau\gg 1$. 
Clearly $\chi_\beta\simeq 0$ implies that no persistent
structure is present whereas, at the other extreme, $\chi_\beta=1$ implies that
all sites are locked in a persistent structure of clusters.

\begin{figure}
\centerline{\psfig{file=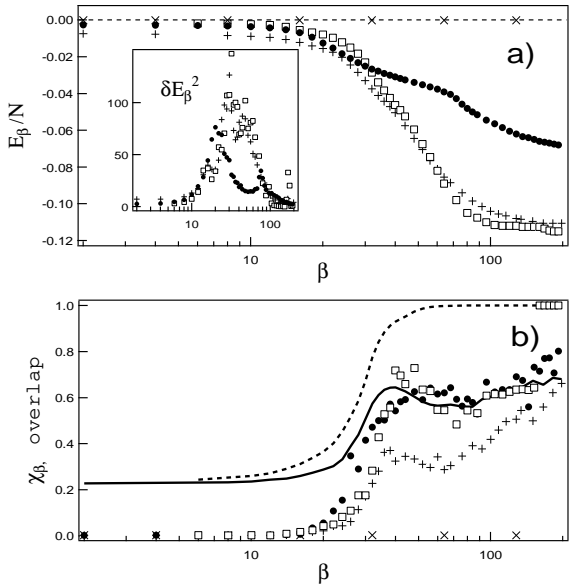,width=8cm}}
\caption{{\bf a)} Energy $E_\beta$ as a function of $\beta$ for random 
($\times$), S\&P500 ($+$) and correlated ($\Box$) data sets of length
$D=1599$ respectively. The results for the S\&P500 data set over the last
$D=400$ days are also shown ($\bullet$). Inset: square energy fluctuation
$\delta E_\beta^2$ vs $\beta$ for the same data sets (same symbols). 
{\bf b)} Autocorrelation $\chi_\beta$ as a function of $\beta$ for the 
same data sets (same symbols). The full (dashed) line refers to the 
overlap with the configuration $s^\star$ for the S\&P500 (correlated)
data set with $D=1599$.}
\label{figeco}
\end{figure}

We monitored these quantities for three different data sets all composed by
$N=443$ time series: {\em 1)} uncorrelated time
series; {\em 2)} the time series of daily returns of the
assets composing the S\&P500 index \cite{notanorma,Mantegna&Stanley} {\em
3)} correlated time series generated by Eq. (\ref{ansatz}) with
given $s_i=s_i^\star$ and $g_s=g_s^\star$. 
The first and the third data sets serve to test the method in cases where we
know the answer. 


Let us start with a truly uncorrelated time series with $D=1599$. 
We compute $C_{i,j}$ and study the corresponding Hamiltonian
$H_c$ by the MC method. We do not expect any clustering to emerge in this 
case. Indeed, the internal energy 
$E_\beta$ stays 
very close to $0$ (see Fig. \ref{figeco}a) for all values of $\beta$
investigated up to $\beta=512$.
Correspondingly no persistent cluster arises, i.e. $\chi_\beta\simeq 0$. 

The results change turning to correlated data. Let us first discuss 
the S\&P500 data for $D=1599$: As Fig. \ref{figeco}a shows,
for $\beta\approx 20$ the energy $E_\beta$ starts 
deviating significantly from zero. For $\beta>20$ persistent clusters are 
present: $\chi_\beta$ rapidly
raises from zero and it has a maximum at $\beta\approx 40$ (see Fig.
\ref{figeco}b). The energy fluctuations reported in the inset 
shows a broad peak of intensity marking the onset of an ordered low 
temperature phase. As $\beta$ increases the
dynamics is significantly slowed down. At
$\beta\approx 200$ the energy reaches a minimal value $E_\beta\simeq -0.11 N$
and does not decrease significantly increasing $\beta$ at least up to
$\beta=4095$. This energy is smaller than that of the ferromagnetic state 
($E_f=-0.086 N$), with all sets in the same cluster. The system in
this range of temperatures visits only few configurations. 

\begin{figure}
\centerline{\psfig{file=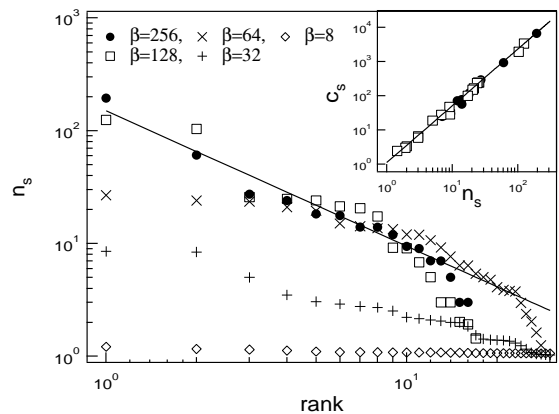,width=8cm}}
\caption{Rank plot of $n_s$ for several values of $\beta$. The line
corresponds to $n\sim {\rm rank}^{-1.2}$.Inset:
$c_s$ versus $n_s$ for $\beta=256$ ($\bullet$) and $\beta$ ($\Box$).
The line corresponds to  $c/sim n^{1.66}$.}
\label{figcls}
\end{figure}

The statistical properties of cluster configurations, as $\beta$ varies, are
shown in Fig. \ref{figcls}. For small $\beta$ only small clusters survive to
thermal fluctuations. As $\beta$ increases a distribution of cluster sizes
develops. At low temperatures the rank order plot of $n_s$ reveals a broad
distribution of clusters with the largest aggregating more than $190$ sets.
By a power law fit of this distribution, we find that the number of clusters 
with more than $n$ sets decays as $n^{-0.83}$. The scatter plot of $c_s$
versus  $n_s$ also reveals a non-trivial power law dependence
$c_s\sim n_s^{1.66}$. This gives a statistical characterization of
the dominant configurations of clusters at low energy\cite{notaecon}.

For $D=400$ we find two transitions at $\beta_1\approx 20$ and at 
$\beta_2\approx 80$ which are signalled by bending in the $E_\beta$ curve
and by peaks in the $\delta E_\beta^2$ vs $\beta$ plot. At the first
temperature clusters start to appear. For $\beta<\beta_2$ the largest
cluster groups less than $30$ sets and for $\beta>\beta_2$ 
larger clusters $n_s\approx 100$ appear. This hints at a time dependence 
of correlations, which are averaged in the $D=1599$ data set.
For even shorter time series we found that sampling errors, acting like a 
temperature, destroy large clusters and only relatively small clusters 
($n_s<40$ for $D=60$) were found. 


We build a syntetic correlated data set of $D=1599$ points 
using Eq. (\ref{ansatz}) with
a structure ${\cal S}={\cal S}^\star$ and parameters 
${\cal G}={\cal G}^\star$. The structure ${\cal S}^\star$ is a typical
low energy configuration for the S\&P500 data set 
for $D=1599$.
 The parameters $g_s^\star$ where deduced from the $n_s$ and $c_s$ 
of this configuration, via Eq. (\ref{xhat}). 
The distribution of eigenvalues of 
$C_{i,j}$ is shown in Fig. \ref{figmat}a. This data set
is useful for at least two reasons: first it allows one to understand to what
extent a struture of correlation put by hand with the form dictated
by Eq. (\ref{ansatz}) can be correctly recovered. 
Secondly it allows us to compare the results found for the S\&P500 data with 
those of a time series with correlations described by Eq. (\ref{ansatz}),
of a similar nature.

For $\beta<150$, 
the behaviors of $E_\beta$, $\delta E_\beta^2$ and $\chi_\beta$ 
are similar to those found for the S\&P500 data 
(see Fig. \ref{figeco}). A second, sharp peak in $\delta E_\beta^2$ at 
$\beta\approx 170$ 
signals a new clustering phase transition. Below this temperature, as shown 
by the plot of $\chi_\beta$ (fig. \ref{figeco}b), the MC dynamics freezes into
the original structure ${\cal S}^\star$. 
The overlap with the configuration ${\cal S}^\star$, defined as in 
Eq. (\ref{chit}) as the fraction of ``bonds'' $s_i=s_j$ for which 
$s_i^\star=s_j^\star$, quickly converges to $1$ (see fig. 
\ref{figeco}b) for the syntetic time series, 
whereas it remains around $60\%$ for the S\&P500 data set. 
This, on one hand means that the original structure ${\cal S}^\star$ 
can be recovered quite efficiently. On the other hand, 
it suggests that several cluster configuration compete at low temperatures 
in the S\&P500 data set. 

Eq. (\ref{ansatz}) with a single cluster configuration 
($\beta\to\infty$), is inadequate to 
capture the full complexity of the correlations in the S\&P500 data set. 
Probabilistic clustering, where several cluster structures 
${\cal S}$ are allowed with their Gibbs probability $P\{{\cal S}\}$
(and finite $\beta$) provides a much better approximation.
At finite $\beta$ each set $i$ may belong to several clusters and we
can measure the corresponding coupling strenght 
$g_{s,i}(\beta)=\langle{\hat g_s \delta_{s,s_i}}\rangle_\beta$.
Taking these as the parameters of the generalized model
\begin{equation}\FL
\xi_i(d)=\frac{\sum_s\sqrt{g_{s,i}(\beta)}\eta_s(d)+\epsilon_i(d)}
{\sqrt{1+\sum_s g_{s,i}(\beta)}},
\label{fitbeta}
\end{equation}
we can build an artificial time series $\vec\xi_i$ and compute 
the correlation matrix $C_{i,j}^{(\beta)}(D)$.
Here $\beta$ is a free parameter which can be adjusted to ``fit'' 
the S\&P500 correlation matrix.  
The eigenvalue spectra of the two matrices are compared in fig. \ref{figmat}b
for $\beta=48$.  The value of $\beta$ was choosen by 
visual inspection as that giving the best fit. The curves are remarkably
close, suggesting that Eq. (\ref{fitbeta}) provides a good
statistical description of the correlations among assets.
Fig. \ref{figmat}b also shows the {\em noise undressed} matrix 
$C_{i,j}^{(\beta)}(\infty)$, which allows one to appreciate the effect of 
noise dressing. As expected, noise mainly affects small eigenvalues.

The applicability of the method can be extended considerably to a generic
data set $\{\vec x_i\}_{i=1}^N$. $\vec x_i$ need not be a time series. 
The distribution of $x_i(d)$ need not be Gaussian and it does not even need 
to be the same across $i$. For example, $x_i(d)$ may be the measure of the 
$d^{\rm th}$ feature of the $i^{\rm th}$ 
object or the concentration of species $i$ in  the $d^{\rm
th}$ sample of an experiment. 
The idea is to map the data set $\vec x_i$ into a Gaussian time series 
$\vec\xi_i$ to which we apply Eq. (\ref{ansatz}). 
The mapping results from requiring
that non-parametric cross-correlations $\tau_{i,j}^x=\tau_{i,j}^\xi$ 
are preserved. To do this in practice we compute Kendall's $\tau$
\cite{Kendall} for the $\vec x_i$ data sets: $\tau_{i,j}^{x}
=\langle{\hbox{sign} [x_i(d)-x_i(d')]\hbox{sign} 
[x_j(d)-x_j(d')]}\rangle_{d<d'}$. 
We note that, for two infinite Gaussian time series
with correlation $c$ we have $\tau=\frac{2}{\pi}[\tan^{-1}
\sqrt{\frac{1+c}{1-c}}-\tan^{-1}\sqrt{\frac{1-c}{1+c}}]$. 
Inverting this relation, we find the correlation $c=C_{i,j}$ as a function
of $\tau=\tau^\xi_{i,j}=\tau^x_{i,j}$. 
This allows us to build the Hamiltonian which can then be studied.

With respect to ref. \cite{Domany}, our approach does not need any 
assumption on the form of the Hamiltonian. 
As input, the method only needs the correlation matrix $C_{i,j}$ (or 
$\tau_{i,j}$). The range of interactions is set by the correlations themselves.
For small $D$, the local interaction of ref. \cite{Domany} may well be more 
efficient in capturing the structure of data. Our method is most useful in
cases where $D\sim N\gg 1$. These ideas can clearly
be extended to models of correlations different from Eq. (\ref{ansatz}).

I acknowledge R. Zecchina, R. Pastor-Satorras and L. Giada for interesting 
discussions and R. N. Mantegna for providing the S\&P500 data.


\begin{references}
\bibitem{notaecon} From the conomic point of view, we find reasonable results:
companies in the same sector belong to the same cluster. We shall not discuss
these issue here, but rather focus on the clustering method.
\end{references}
\end{document}